\pdfoutput=1

\documentclass[12pt]{article}

\usepackage{amssymb}
\usepackage{array}
\usepackage{graphicx}
\usepackage{amsmath}

\newcommand{\be}{\begin{equation}}
\newcommand{\ee}{\end{equation}}

\newcommand\eqnn[2]{
        \begin{equation}#2\label{#1}
        \end{equation}}
\newcommand\elnn[2]{
        \begin{eqnarray}
        #2\label{#1}
        \end{eqnarray}}
\newcommand\elnnsingle[2]{
        \begin{align}
        \begin{split}
        #2\label{#1}
        \end{split}
        \end{align}
}
\newcommand\elnnn[1]{
        \begin{eqnarray}
        #1
        \end{eqnarray}}

\begin{document}

\markboth{J.~D. Carrick and F.~I. Cooperstock }
{General relativistic dynamics applied to the rotation curves of galaxies
}


\title{
         General relativistic dynamics applied to the rotation curves of galaxies
}

\begin{center}
  J.~D. Carrick and F.~I. Cooperstock\\
{\small \it Department of Physics and Astronomy, University 
of Victoria} \\
{\small \it P.O. Box 3055, Victoria, B.C. V8W 3P6 (Canada)}\\

{\small \it e-mail addresses: cooperst@uvic.ca, jcarrick@uvic.ca}
\end{center} 



\maketitle

\begin{abstract}

We extend our general relativistic analysis of galactic rotation curves with
galaxies NGC 2841, NGC 2903 and NGC 5033. As before, we employ the solution of the
Einstein field equations of general relativity with an expansion in Bessel
functions. As in our earlier studies, the fits to the data are found to be very
precise and the calculated baryonic masses are lower than those based upon Newtonian
gravity.
Also as in our previous studies, the galactic radii at which the optical
luminosities terminate are seen to correlate
with densities near $10^{-21.75}$ kg$\cdot$m$^{-3}$. This concordance lends further
support to the correctness of the procedure as well as providing a potentially
valuable piece of information in the understanding of galactic evolution.

\end{abstract}





\section{Introduction}

It is widely believed that Einstein's theory of gravity is primarily a theory for
very strong gravitational fields and that it provides only very small corrections to
Newtonian predictions for weak fields. Indeed this is very often seen to be the case
as in planetary motion studies. Since the gravitational fields within galaxies are
generally weak and the velocities of the component stars are non-relativistic, general
relativity was never brought into the analysis prior to our work in \cite{CT},
\cite{CT2} and \cite{CT3}.  
What we had found was that general relativity does indeed yield results that differ
from
those derived from Newtonian gravity for \textit{extended} gravitationally bound
systems, even for such weak-field systems. From the relatively flat galactic
rotation curves that were first observed
by V. Rubin, Newtonian gravity yielded only one possible conclusion: the galaxies
must be encircled by vast spherical halos of ``dark matter'' with masses up to an
order of magnitude larger than those indicated by their visible contents. Even
earlier, F. Zwicky had proposed that vast quantities of dark matter were required to
propel the galaxies in galactic clusters such as in the Coma Cluster. While
there have been several ideas proposed regarding the physical constitution of this
mysterious matter, no direct detection has ever been achieved. Moreover, while
probably the majority of investigators are convinced of its reality, a variety of
researchers have come to doubt its very existence. In recent years, the dark matter
issue has evolved into one of the most important problems in theoretical physics,
astrophysics and cosmology.

Our own attitude is guided by ``Occam's razor''. Unless such a new element (and a
very bizarre one at that) is seen
to be absolutely necessary, we prefer to work within the framework of its presumed
absence. An essential point is this: \textit{the known data from galactic rotation
curves can be accommodated with at most relatively little extra matter when the
analysis is performed with Einstein's as opposed to Newton's gravity and it is
almost universally believed that Einstein's general relativity is our best theory of
gravity.} However, it must be stressed that Einstein's theory is much richer than
Newton's theory. General relativity can also accommodate large halos of extra
encircling matter and we have determined the criterion for deducing the extent,
if any, of the presence of extra matter \cite{CT4} \cite{C1}. It is determined by
the degree of velocity dispersion for stars that lie above and below the galactic
symmetry plane; the greater the degree of velocity dispersion that is indicated, the
lesser the degree of indicated extra matter. Hopefully such dispersion data will
eventually be available in sufficient quantity and quality to make such a
determination possible. 

In the meantime, we have continued our analysis of rotation curves, again fitting
the known data with the appropriate solutions of the Einstein field equations. We
see the pattern continuing of smaller indicated galactic masses than those
calculated on the basis of Newtonian theory. As well, we see a continuation of the
correlation of the onset of optical luminosities of the new galaxies studied at the
same approximate value of average density, namely $10^{-21.75}$ kg$\cdot$m$^{-3}$.
It is
remarkable that Einstein's theory has within it the power to reveal such a trend
solely on the basis of rotation curve data. It is significant that the optical
thresholds occur at quite a variety of distances from the galactic axes of rotation
so it is not as if we are dealing with a  ``cookie cutter'' situation. This
confluence of results fortifies the trust in the methodology.

While various researchers are now turning to gravitational lensing in the search for
evidence for dark matter, probably the majority of researchers regard the flat
galactic rotation curves as the key indicators. Some have indicated that it is
evidence garnered from microwave early-universe observations that is most compelling
but there are problems associated with the dark matter theory of galaxy formation.
For example, this theory permits 10 to 100 times as many small galaxies as those
which are actually observed. 

Clearly the issue is 
of paramount importance given that dark matter is said to comprise the dominant
constituent of an extended galactic mass by multiple factors 
 The dark
matter enigma has influenced particle theorists to devise acceptable candidates for
its constitution.  At the present time, various researchers are
hoping to discover dark matter particles in the LHC (Large Hadron Collider)
experiments at CERN in Geneva. While physicists and astrophysicists have pondered
the dark matter issue, other researchers have devised new theories of gravity to
account for the observations
(see for example \cite{bek1,bek3,bek4,bek5}). However the latter approaches, however
imaginative, have met with understandable skepticism, as they have been devised
solely for the purpose of the task at hand.
General relativity was the existing preferred theory of gravity long before dark
matter was even contemplated and it remains the preferred theory to this day. 
General relativity 
has been successful in every test that it has encountered, going 
beyond Newtonian theory where required. Therefore, should it actually transcend
Newtonian gravity in resolving the galactic rotation curve issue, it would greatly
alter the
understanding of some basic aspects in physics and astrophysics.

In dismissing general relativity in favour of Newtonian gravitational theory for the
study of galactic 
dynamics, insufficient attention has been paid to the fact that the 
stars that compose the galaxies are essentially in motion under 
gravity alone (``gravitationally bound''). It had been known for many years, in fact
since 
the time of Eddington, that the gravitationally bound problem in 
general relativity is an intrinsically non-linear problem 
even when the conditions are such that the field is weak and the 
motions are non-relativistic, at least in the time-dependent case. 
\textit{Most significantly, we have found that under these 
conditions, the general relativistic analysis of the problem is also 
non-linear for the stationary (non-time-dependent) case at hand.} 
Thus the intrinsically linear Newtonian-based approach used prior to our work has
been inadequate for the description of the galactic 
dynamics and Einstein's general relativity must be brought into the 
analysis within the framework of established gravitational 
theory.
This is an essential departure from conventional thinking on the 
subject and it leads to major consequences.

Since our initial posting \cite{CT}, many colleagues have offered 
their comments and criticisms. To provide some perspective, we briefly address the
most common areas of contention in Section 5.  
The totality of the issues known to us have been discussed in \cite{CT2},
\cite{CT3}, \cite{CT4} and \cite{C1}.  
In those papers, we developed the theory in more detail and applied it to three
galaxies and the Milky Way. We also developed
a new observational discriminator 
for assessing the degree, if any, of external matter that may lie 
beyond the visible/HI regions. This is determined by examining the galactic rotation
curves
at different galactic latitudes, bringing into consideration the \textit{global}
dynamical
structure of the galaxy. It is well to emphasize that the demand for global
consistency applies not only to our own but also to all proposed models and
theories. 


\section{Axially-Symmetric Model Galaxy}

In the interests of self-containment, we set out the essentials of the general
relativistic mathematical structure for the problem. 
When we consider the complexity of the detailed structure of a typical 
galaxy with its arms and irregular density variations, it becomes 
clear that the modeling within the context of the complicated theory 
of general relativity demands some simplifications, at least at this point in our
facility in handling this rich and demanding theory. As long as 
the essence of the structure is captured, these simplifications are 
justified and we can derive valuable information. 
To capture the essence, we consider an idealized model that lends itself to detailed
analysis, a time-independent uniformly rotating fluid without pressure and symmetric
about its 
axis of rotation. 
In generality, the stationary axially symmetric metric in consort with the model can
be described in the form
\eqnn{Eq1}{
        ds^2 =        -e^{\nu-w}( udz^2+dr^2)
                -r^2 e^{-w} d\phi^2+e^w(cdt-Nd\phi)^2
}
where $u$, $\nu$, $w$ and $N$ are functions of cylindrical polar 
coordinates $r$, $z$. It is easy to show that to the order required, 
$u$ can be taken to be unity. 
It is simplest to work in the frame that is comoving with the
matter,
\eqnn{Eq2}{
        U^i = {\delta}_0^i
}
where $U^i$ is the four-velocity.
        This is reminiscent of the standard approach that is
        followed for FRW (Friedmann-Robertson-Walker) cosmologies. However, the FRW
spacetimes
        are homogeneous and they are not stationary.
The comoving approach was taken in the pioneering paper by van Stockum \cite{vs} who 
set $w=0$ from the outset.
        Interestingly, the geodesic equations imply that
        $w=constant$ (which can be taken to be zero as in
        \cite{vs}) even for the \textit{exact} Einstein field
        equations as studied in \cite{vs}. In fact the
        requirement that $w=0$ can be seen directly using
        (\ref{Eq2}) and the metric equation $g_{ik}U^iU^k=1$
        \cite{CT2}. 
As we discussed in 
our previous work, we perform
a purely \textit{local} ($r,z$ held fixed at each point when
taking differentials) transformation.
It is to be noted that this local transformation is used only to deduce the
connection between $N$ and $\omega$ (and hence $V$). All subsequent work continues
in the original unbarred comoving frame.
The localized transformation is 
\eqnn{Eq3}{
        \bar{\phi} = \phi + \omega(r,z)\,t
}
which is adjusted to locally diagonalize the metric. In this way, we are able to 
determine the local angular velocity $\omega$ and the tangential 
velocity $V$ as
\elnn{Eq4}{
        &\omega
        = \frac{Nce^w}{r^2e^{-w}-N^2e^w}
        \approx \frac{Nc}{r^2} \label{Eq4top}\\
      & V      
      =\omega r
}
with the approximate velocity value $Nc/r$ applicable for the weak fields under 
consideration. The Einstein field equations to order $G^1$
with $w$ kept for later comparison, are
\elnnsingle{Eq5top}{
        2r\nu_r+ N_r^2-N_z^2 =0, \\
        r\nu_z +N_r N_z =0,  \\
        N_r^2 + N_z^2 +2r^2(\nu_{rr}+\nu_{zz}) =0, \\
        N_{rr} +N_{zz} - \frac{N_r}{r}=0,
}
\elnn{Eq5}{
        \left(w_{rr} +w_{zz} +\frac{w_r}{r}\right)
        + \frac{3}{4}r^{-2} (N_r^2 + N_z^2)& \nonumber \\
        + \frac{N}{r^2}\left(N_{rr} +N_{zz} -\frac{N_r}{r}\right)
        - \frac{1}{2}(\nu_{rr}+\nu_{zz})
        &= 8{\pi}G\rho/ c^2
}
where $G$ is the gravitational constant and $\rho$ is the mass 
density. Subscripts indicate partial differentiation with respect to 
the indicated variable. Fortunately, these rather complicated equations are easily
combined to give
\eqnn{Eq5a}{
        \nabla^2 w +\frac{N_r^2+N_z^2}{r^2}=\frac{8\pi G\rho}{c^2}
}
where the first term is the flat-space Laplacian in cylindrical polar coordinates
\eqnn{Eq5b}{
        \nabla^2 w \equiv w_{rr} + w_{zz} + \frac{w_r}{r}
}
and $\nu$ is determined by simple integration.

>From the application of the freely gravitating constraint (i.e. stress-free motion)
and the requirement that $w=0$ which arises from the choice of comoving coordinates,
the field equations 
for $N$ and $\rho$ in this globally dust distribution are reduced to 
\elnnn{
        N_{rr} + N_{zz} - \frac{N_r}{r} =0
        \label{Eq9a} \\
        \frac{N_r^2 + N_z^2}{r^2} = \frac{8{\pi}G\rho}{c^2}.
        \label{Eq9b}
}
Note that with the minus sign in (\ref{Eq9a}), $N$ does \textit{not}
        satisfy the Laplace equation.
Note also that from both the field equation for $\rho$ and the expression 
for $\omega$ that $N$ is of order $G^{1/2}$. This is a point that has been
misunderstood by some of our critics, leading them to erroneous conclusions. (We
discuss this briefly later in the paper.)

The non-linearity of 
the galactic dynamical problem is evident through the
non-linear relation between the functions $\rho$ and $N$.
        While we have eliminated $w$ either by using the geodesic
        equations to get (\ref{Eq9b}) or by the metric equation and
        the choice of comoving coordinates, $N$ cannot be eliminated
        and hence non-linearity is intrinsic to the study of the
        galactic dynamics.
 Rotation under freely 
gravitating motion is the key factor at play in the present problem. 

It is worthy of note that (\ref{Eq9a}) can be expressed as
\eqnn{Eq10}{
        \nabla^2\Phi =0
}
where
\eqnn{Eq10a}{
        \Phi \equiv \int\frac{N}{r}dr.
}
Thus, flat-space harmonic functions $\Phi$ are the generators of 
the axially symmetric stationary pressure-free weak fields that we
seek.
        (In fact Winicour \cite{Winicour} has shown that all such sources,
        even when the fields are strong, are generated by such flat-space
        harmonic functions.)
We refer to these as ``generating potentials''. It is noteworthy that these
generating potentials play a different role in general relativity than do the
ordinary potentials of Newtonian gravitational theory even though both functions are
harmonic. 
Using (\ref{Eq4}) and (\ref{Eq10a}), we have the expression for 
the tangential velocity of the distribution,
\eqnn{Eq11}{
        V=c\frac{N}{r} =c\frac{\partial {\Phi}}{\partial{r}}.
}
In Newtonian gravity, the potential gradients relate to acceleration whereas here,
they relate to velocity. These potentials do not play the same role in the two
theories of gravity.

We now have the necessary mathematical framework in place.

\section{Connecting the Observed Galactic Rotation Curves to the Model}

We first consider the ideal strategy for galactic modeling, given the nature of the
equations at hand.
Since the field equation for $\rho$ is non-linear, the simpler way to 
proceed 
is to first find the required generating 
potential $\Phi$ and from this, derive an appropriate function $N$ 
for the galaxy that is being analyzed. With $N$ found, (\ref{Eq9b}) 
yields the density distribution. If this agrees with the 
observations, the efficacy of the approach is established. This is in 
the reverse order of the standard approach to solving gravitational 
problems but it is most efficient in this formalism because of the 
existence of one linear field equation.

For any given galaxy, a suitable set of composing functions for the series that
satisfies the linear equation is required. Once found, this yields the generating
potential. 
With cylindrical polar 
coordinates, it is simplest to use separation of variables leading to the following
solution 
for $\Phi$ in (\ref{Eq10}):
\eqnn{Eq12}{
        \Phi = Ce^{-k\mid z \mid}J_0(kr)
}
where $J_0$ is the Bessel function $m=0$ of Bessel $J_m(kr)$ and $C$
is an arbitrary constant 
        (see for example \cite{AW}). Using this form of solution, the 
        absolute value of $z$ must be used to provide the proper 
        reflection of the distribution for negative $z$. While this 
        produces a discontinuity in $N_z$ at $z=0$, it is important 
        to note that in the problem at hand, this discontinuity is 
        consistent with the general case of having a density 
        z-gradient discontinuity at the plane of reflection symmetry. 
        This point has been the subject of considerable attention in the literature. We
will return to this issue later.

The beauty of a linear equation is that it allows for linear superposition of
solutions. 
>From (\ref{Eq10}) we express the general
solution of this form as the linear superposition
\eqnn{Eq13}{
        \Phi = \sum_{n}C_ne^{-k_n |z|}J_0(k_nr).
}
We choose integer $n$ as required to the level of accuracy that we wish to achieve.
>From (\ref{Eq13}) and (\ref{Eq11}), the tangential
velocity is
\eqnn{Eq14}{
        V= -c\sum_{n} k_n C_n e^{-k_n |z|}J_1(k_nr).
}
For this, we have used the Bessel relation $dJ_0(x)/dx= - J_1(x)$ (see, e.g.
\cite{ford}). From (\ref{Eq11}), we see that if $N$ should exceed $r$, the velocity
would exceed $c$. This does not occur because with our choice of separable
solutions, the velocity is given by (\ref{Eq14}). As $r$ approaches $0$, this
function falls as $r^1$ (i.e. $N$ approaches $r=0$ as $r^2$) and so $V$ falls to $0$
as we see as well in the plots of the rotation curves. 
Thus the potential problem referred to by Wiltshire \cite{Wilt} is not present in
our case. As well, for large r, the Bessel functions fall as $1/\sqrt{r}$ and hence
the velocity goes to $0$ for large $r$. In general, this would still lead to a large
amount of matter external to the galaxy. However, by our selecting the right
multiplying coefficients in the Bessel solution sequence, we can achieve consistency
with a very limited amount of external matter, far less than that which is indicated
on the basis of Newtonian gravity.  We have shown this by fitting fictitious
faster-than-$1/\sqrt{r}$ fall-off data for the extension of the observed points in
the velocity profile (see \cite{CT4}). 

We choose the $k_n$ so that the $J_0(k_nr)$ terms are orthogonal to each other.
The Bessel functions $J_0(kr)$ satisfy the 
        orthogonality relation: 
\eqnn{Eq14a}{
\int_{0}^{1} J_0(k_nr)J_0(k_mr)rdr 
        \propto \delta_{mn} 
}
where $k_n$ are the zeros of $J_0$ at 
        the limits of integration. 
With only 10 functions with parameters 
$C_n$, $n\in\{1\dots 10\}$, we have been able to achieve an excellent
fit
to the velocity curve 
for the Milky Way  
and for NGC 3031, NGC 3198 and NGC 7331 (see \cite{CT4}). In this paper, we advance
beyond what we achieved in our earlier work. 

        It should be noted that the $J_1(x)$ Bessel functions are $0$ at $x=0$ 
        and oscillate with decreasing amplitude, falling as 
        $1/\sqrt{x}$ asymptotically \cite{ford}.  
        However, this alone does not assure a realistic 
        fall-off of matter. We have dealt at length with this issue in \cite{CT4}
(see also \cite{C1}). 
        Also, our curves drop as $r$ approaches $0$.  This 
        is in contrast to the Mestel \cite{mes} and MOND
        \cite{bek1,bek3,bek4} curves that are flat
        everywhere. 

 
>From (\ref{Eq11}) and (\ref{Eq14}), the $N$ function is determined 
in detail and from (\ref{Eq9b}), the density distribution follows. \textit{Most
significantly, our correlation 
of the flat velocity curve is achieved with disk mass up to an order of 
magnitude smaller than the halo mass of dark 
matter proposed by earlier studies.} (See e.g.\cite{clew1,clew2}).

It is to be emphasized that general relativity does not distinguish between the
luminous and 
non-luminous contributions. The deduced $\rho$ density distribution 
is derived from the totality of the two. Any substantial amount of 
ordinary baryonic non-luminous matter  
would necessarily lie in the flattened region 
relatively close to $z=0$ because this is the region of significant 
density $\rho$ and would be due to dead stars, planets, neutron stars and 
other normal non-luminous baryonic matter debris. Each term within 
the series has $z$-dependence of the form $e^{-k_n|z|}$ which causes the steep
density fall-off profile.


This fortifies the picture of a 
standard galactic essentially flattened disk-like shape as 
opposed to a halo sphere. From the evidence provided thus far by 
rotation curves \cite{CT} \cite{CT2} \cite{CT3}, there is no support for the widely
accepted notion of the 
\textit{necessity} for massive halos of dark matter surrounding visible 
galactic disks: conventional gravitational theory, namely general 
relativity, can account for the observed flat galactic rotation 
curves linked to essentially flattened disks with no evident need for 
dark matter, at least not with the velocity distribution data presently at
our disposal.

We recall from our earlier work, general relativity in conjunction with the figures
provided by Kent 
\cite{Kent} for optical intensity curves and our log density profiles 
for three galaxies NGC 3031, NGC 3198 and NGC 7331 provide a common thread: all three 
galaxies indicate the threshold 
density for the onset of visible galactic light as we probed in the 
radial direction at approximately $10^{-21.75}$ kg$\cdot$m$^{-3}$. 
%
This led us to hypothesize that this density is the universal optical 
luminosity threshold for galaxies as tracked in the radial direction. 
In what follows, we will present further evidence in support of this hypothesis from
the investigation of three additional galaxies. 

As well, the radius
at which the optical luminosity fall-off occurs can be 
predicted for other sources using this particular density parameter. The 
predicted optical luminosity fall-off for the Milky Way is at a 
radius of 19-21 Kpc based upon the density threshold indicator that 
we have determined. It will be interesting to test this indicated range according to
our hypothesis with more refined
data that the astronomers will be able to accumulate in the future. 

\section{Analysis of Galaxies NGC 2841, NGC 2903 and NGC 5033}
Velocity curve-fits and density profiles (see figures) for galaxies NGC 2841, NGC
2903 and NGC
5033 were obtained using the methods outlined in Section 3. As in \cite{CT4}, only
ten parameters were used per data set, the values for which were obtained via
least-squares minimization. (Data tables follow below.) The value for the radius at
which the visible light is first
observed was obtained from Kent \cite{Kent} for galaxies NGC 2903 and NGC 5033, and
was obtained from Macri et al \cite{Macri} for NGC 2841.

\begin{figure}
\begin{center}
\includegraphics[width=5.00in]{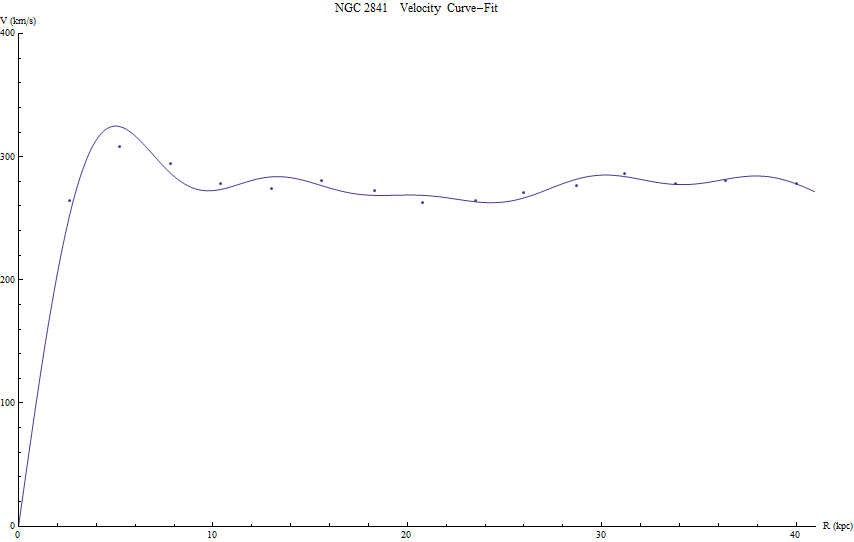}
\vskip 0.4 in
\includegraphics[width=5.00in]{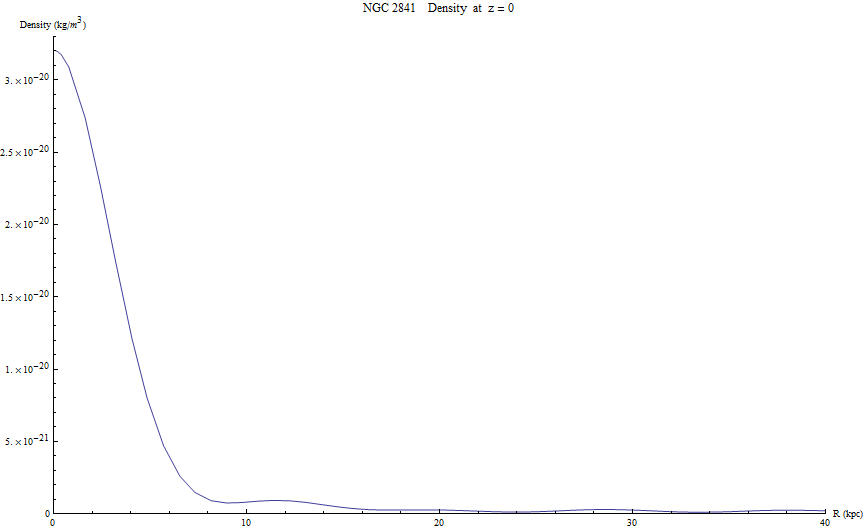}
\end{center}
\caption{
\label{fig:ngc2841}
Velocity curve-fit and derived density for NGC 2841}
\end{figure}

\begin{figure}
\begin{center}
\includegraphics[width=5.00in]{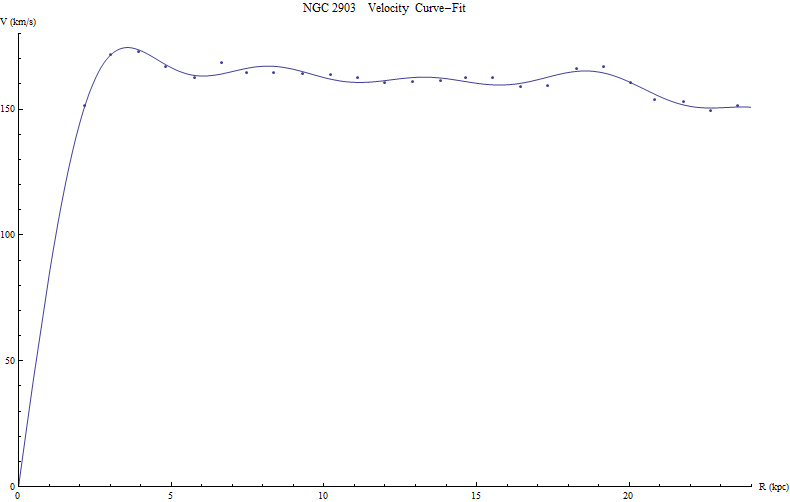}
\vskip 0.4 in
\includegraphics[width=5.00in]{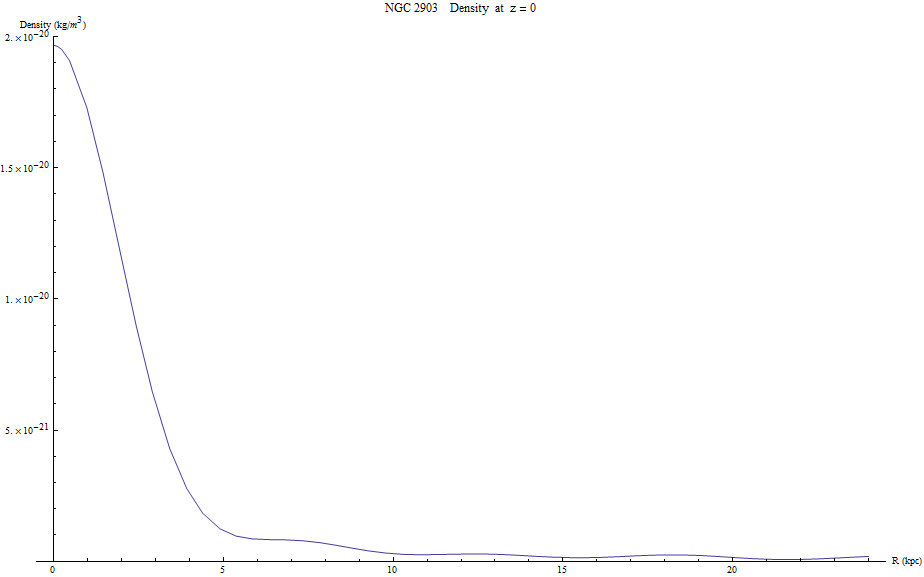}
\end{center}
\caption{
\label{fig:ngc2903}
Velocity curve-fit and derived density for NGC 2903}
\end{figure}

\begin{figure}
\begin{center}
\includegraphics[width=5.00in]{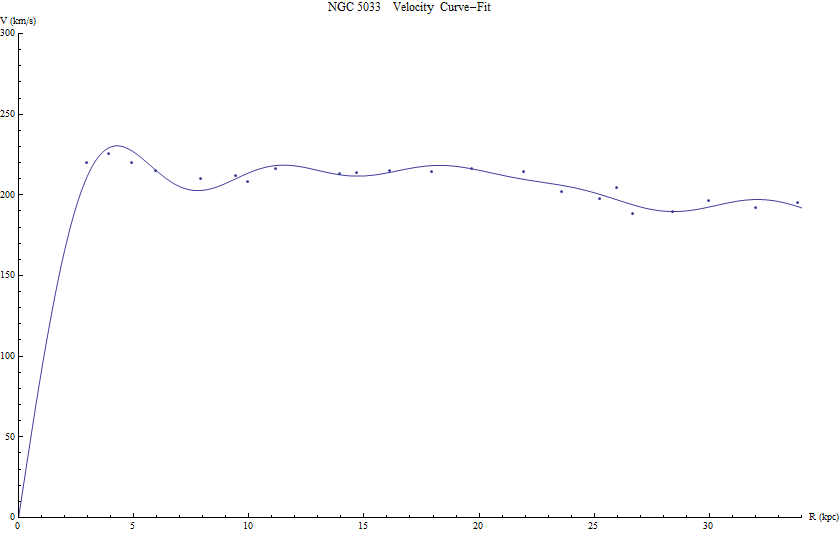}
\vskip 0.4 in
\includegraphics[width=5.00in]{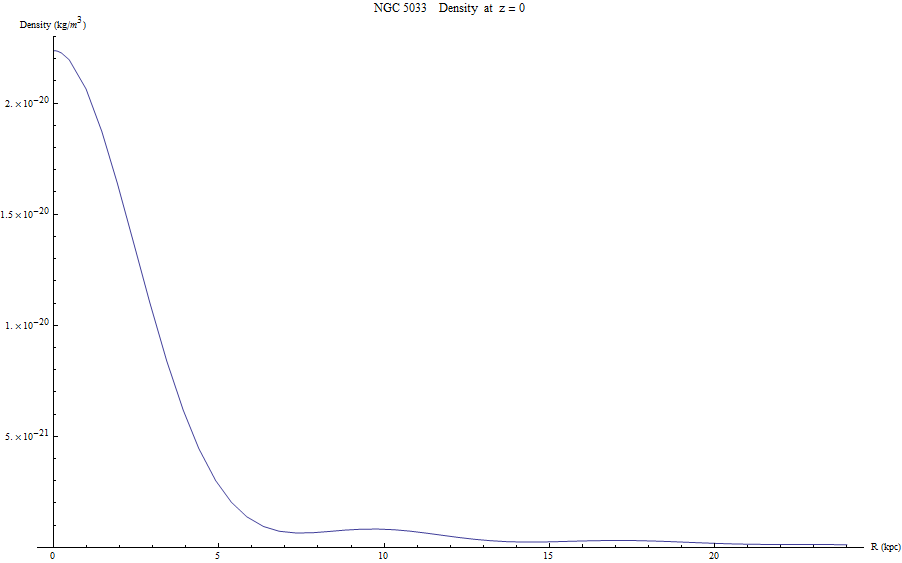}
\end{center}
\caption{
\label{fig:ngc5033}
Velocity curve-fit and derived density for NGC 5033}
\end{figure}

\begin{table}[h!]
\centering
\setlength{\extrarowheight}{2pt}
\begin{tabular}{ lr}
\hline
$-C_n*k_n$ & $k_n$ (Kpc$^{-1}$) \\
\hline
0.001263951207 & 0.06680070994\\
0.0003509253903 & 0.1533355031\\
0.0002902041947 & 0.2403813309\\
0.0001233680182 & 0.3275426233\\
0.0001966403398 & 0.4147477141\\
0.0001053122057 & 0.5019739991\\
0.00009540622257 & 0.5892121286\\
0.0001110222640 & 0.6764575425\\
0.00006439084160 & 0.7637077537\\
0.00007895785801 & 0.8509612908\\
\hline
\end{tabular}
\caption{Curve-fitted coefficients for NGC 2841}
\label{tab:}
\end{table}

\begin{table}[h!]
\centering
\setlength{\extrarowheight}{2pt}
\begin{tabular}{ lr}
\hline
$-C_n*k_n$ & $k_n$ (Kpc$^{-1}$) \\
\hline
0.0009976025857 & 0.09619302231\\
0.0002429098737 & 0.2208031244\\
0.0002257927027 & 0.3461491165\\
0.0001463861599 & 0.4716613776\\
0.0001087456698 & 0.5972367083\\
0.00007440663732 & 0.7228425587\\
0.00008677052850 & 0.8484654652\\
0.00002425953621 & 0.9740988612\\
0.00006760519096 & 1.099739165\\
3.255331982$*10^{-6}$ & 1.225384259\\
\hline
\end{tabular}
\caption{Curve-fitted coefficients for NGC 2903}
\label{tab:}
\end{table}

\begin{table}[h!]
\centering
\setlength{\extrarowheight}{2pt}
\begin{tabular}{ lr}
\hline
$-C_n*k_n$ & $k_n$ (Kpc$^{-1}$) \\
\hline
0.001263951207 & 0.06680070994\\
0.0003509253903 & 0.1533355031\\
0.0002902041947 & 0.2403813309\\
0.0001233680182 & 0.3275426233\\
0.0001966403398 & 0.4147477141\\
0.0001053122057 & 0.5019739991\\
0.00009540622257 & 0.5892121286\\
0.0001110222640 & 0.6764575425\\
0.00006439084160 & 0.7637077537\\
0.00007895785801 & 0.8509612908\\
\hline
\end{tabular}
\caption{Curve-fitted coefficients for NGC 5033}
\label{tab:}
\end{table}

\begin{table}[h!]
\centering
\setlength{\extrarowheight}{2pt}
\begin{tabular}{ lcccc}
\hline
Galaxy & Critical & Critical & Mass & Kent's Value\\
\null &  Radius (Kpc) & Density ($kg/m^3$) &
($M_{\odot}\cdot 10^{10}$) &
($M_{\odot}\cdot 10^{10}$)\\
\hline
NGC 2841 & 17.0 & 10$^{-21.60}$ & 51.5 & 74.6 \\
NGC 2903 & 12.1 & 10$^{-21.59}$ & 8.9 & 19.4 \\
NGC 5033 & 23.2 & 10$^{-21.83}$ & 22.2 & 37.1\\
\hline
\end{tabular}
\caption{Galactic Data}
\label{tab:}
\end{table}

While the visible light profile for NGC 2841 terminates at $r = 17.0$ Kpc, the
HI profile extends to 40 Kpc. The density at $z = 0$, $r = 17.0$ Kpc, i.e. the
critical density, was found to be $10^{-21.60} kg/m^3$.  Integrating through the HI
outer region to $r = 40$ Kpc yields a total mass of $51.5 \times 10^{10}M_\odot$.
This value is considerably less than that obtained by Kent \cite{Kent} of $74.6
\times 10^{10}M_\odot$. For NGC 2903, we found the critical density to be
$10^{-21.59} kg/m^3$ and the mass to be $8.9 \times 10^{10}M_\odot$. Kent
\cite{Kent} found a value of $19.4 \times 10^{10}M_\odot$ for this galaxy, more than
double the model value. Finally,
using rotation curve data from NGC 5033 the value for the critical density was
determined
to be $10^{-21.83} kg/m^3$ and the mass to be $22.2 \times 10^{10}M_\odot$, while
Kent found the mass to be $37.1 \times 10^{10}M_\odot$.

Thus, as before, with nothing more than Einstein's theory of relativity and the
astronomical data of rotation curves, we
have consistently found a value in the range of $10^{-21.6} kg/m^3$ to $10^{-21.8}
kg/m^3$ for the threshold density of the onset of galactic optical luminosity. It is
well to contemplate the sequence of analysis: Firstly, from Doppler shifts of
spectral lines, we deduce the velocities of stars as a function of distance from the
galactic axis of rotation. Secondly, the solution from Einstein's theory of gravity
uses this data to produce a map of galactic averaged matter density as a function of
this distance. Thirdly, we note from astronomical observations the distance at which
the optical luminosity cuts out and from the density plot, we note the density at
which this occurs. This distance is different for each galaxy yet remarkably, the
density in each of now six cases is approximately the same, namely in the
$10^{-21.75} kg/m^3$ range. This remarkable concordance fortifies one's faith in the
correctness of the procedure.

\section{Perspectives}

It is useful to summarize some of the history relating to our work. We had
challenged the traditional view that Newtonian gravity was the adequate theory to
describe the motions of stars in the galaxies, arguing that general relativity was
required within the context of accepted gravitational physics. We had shown that the
known galactic rotation curves could fit into the framework of general relativity
without the demands for vast halos of dark matter surrounding the galaxies. Other
gravity theories had been \textit{specifically designed} to account for the motions
without invoking dark matter. Einstein's theory stands firmly on its own, predating
all such efforts. 

Understandably, our work was problematic for the many researchers who had relied
upon traditional gravity theory for their research. They could ignore the
alternative theories as being outside of their sphere of accepted science but
general relativity was accepted by virtually all researchers as the very best theory
of gravity. Thus our work was the subject of intense scrutiny and criticism. It is
to be emphasized that we have responded to all the issues of criticism and to this
point, as far as we are aware, there has not been any new rebuttal offered by any of
these critics. As well, there has been some work by others lending support to our
findings.

To round out a unified picture of our work, it is worth a brief recount of the key
areas of criticism. The
complete analysis is contained in our earlier papers and in the book \cite{C1}.
A persistent issue 
concerns the nature of the matter 
distribution. Some critics have noted that given the existence of the 
discontinuity of $N_z$ that we had pointed to in \cite{CT}, a 
significant surface tensor $S_i^k$ can be constructed 
whose integral represents an accumulation of mass above and beyond the volume
integral of the regular continuous mass density.

To analyze this claim, we calculated the surface mass that 
was said to be present in the four galaxies that we had first studied in \cite{CT},
\cite{CT2}, \cite{CT3}. This was done by 
integrating 
this supposed surface density.  
For each galaxy, we calculated a numerical value slightly less 
than the mass that we had derived from the volume integral of our 
\textit{continuous} mass density distribution using (\ref{Eq14}), 
(\ref{Eq11}) and (\ref{Eq9b}).  
Upon reflection, we realized  that \textit{the surface integral of this supposed
singular layer is merely a mathematical construct that indirectly 
describes most of the continuously distributed mass by means of the 
Gauss divergence theorem}. The reason that the surface contributions are slightly
less in each case derives from the small contributions that are absent from the
remaining surfaces to bound the entire volume. The totality of bounding surfaces are
required for the application of the Gauss theorem.

The essential point is that singularities have to be interpreted properly. As a
useful example, consider the simple case of a purely vacuum Schwarzschild solution
with mass $m$ for $r>a$ and Minkowski space for $r<a$. 
Since it is a vacuum solution, the only place for the mass to reside is on the
surface at $r=a$. The surface layer construct finds this mass at $r=a$. However in
our present galactic modeling case, the supposed surface layer construct merely
recovers the \textit{volume-distributed} mass in a different manner, i.e. by the
Gauss theorem. By contrast, in the Schwarzschild shell example, there is no volume
distribution of mass present. The surface layer calculation indicates the mass value
of a continuous distribution of matter that could have been squeezed together to
form the shell. Clearly, this example connects directly with our interpretation of
the galactic model singularity. 
Actually, had we followed the standard prescription to calculate mass with the
surface layers, we would have had to ascribe a negative value to it as others have
claimed.
However, as Bondi has discussed in his works, negative mass repels other masses, be
they of positive or negative mass and our careful investigations have shown that
test particles are invariably attracted rather than repelled at positions near the
discontinuous surface. This negates the negative mass surface layer claim.

Our model consists of dust with reflection symmetry about the $z=0$ 
plane. The density naturally increases symmetrically as this plane is 
approached from above and from below with the same absolute value of density
gradient but of 
opposite sign from symmetry. In all generality, the density z-gradient 
will be different from zero as this plane is approached and because 
of reflection symmetry, this gradient will of necessity be 
discontinuous. This is the physical origin of the singularity. In our case, the
singularity is benign, with a ready physical explanation.

Regarding another issue, some researchers misunderstood our work by not appreciating
the effect of rotation on the perturbation sequence.
Typically 
, these authors present the well-known expression of the field equations in 
the harmonic gauge in Cartesian coordinates.
>From the standard description of the 
post-Newtonian perturbation scheme, they conclude that the solution to 
the galactic problem must be the usual Newtonian one and that all 
corrections must be of higher order.  Firstly, we did not use this 
scheme (as noted as well in \cite{maia}).  
We used cylindrical polar coordinates comoving with the 
matter.  For the gravitationally bound system, the metric components are of
\textit{different} orders in $G$. This is a key point that was overlooked by some of
our critics. 
The novel aspect for our problem is 
that the lowest order equation, of order $G^{1/2}$,  
has zero on the RHS and the second equation that would normally be 
the Newtonian Poisson equation, differs in that it has non-linear 
terms.  Thus, the structure of our solution does not proceed as in 
the standard approach 
referred to by our critics. 
In the standard 
approach, the lowest order base solution is the Newtonian solution 
whereas in the galactic problem, the lowest order equation is the 
Laplace equation for which an order $G^{1/2}$ solution is necessary 
and the next order (order $G^1$) equation for the density  
\eqnn{Eq3g}{
        \frac{N_r^2 + N_z^2}{r^2} = \frac{8{\pi}G\rho}{c^2}
}
has non-linear terms in the metric in the form of the squares of the 
derivatives of an order $G^{1/2}$ metric tensor component $N$. Thus, 
our situation is unlike standard iterative perturbation scheme applications.

A key point is that the equations have an inherent non-linearity 
as a result of the fact that the metric components are of different 
orders and the different orders are a necessary consequence of the 
problem being a gravitationally bound one with rotation.

The authors of \cite{BG} arrive at our equations (\ref{Eq5top}),
(\ref{Eq5})
(with $w$ set to zero) apart from the exponential $\nu$ factor which 
they later note can be taken to be a constant scaling factor and find 
the same order of magnitude reduction of galactic mass that we had 
found \cite{CT} starting from their exact solution class.  This 
provides some vindication for our analysis. 

While we have shown that the presently available data on galactic rotation curves
can be explained without vast halos of dark matter surrounding the galaxies, more
complete data above and below the (approximate) galactic symmetry plane could reveal
the presence of more matter than is presently observed.

Clearly it is important to approach the dark matter issue in 
as many ways as possible. After all, from a purely formal point of 
view, general relativity should be able to model vastly extended 
distributions of pressure-free fluids in rotation.  In this vein, we 
have constructed a test in principle that relies upon data in the 
\textit{visible/HI} regime thus making it particularly useful. When we examine
the different possible fall-offs as in \cite{CT3}, 
we see that different profiles beyond
the HI region imply different mass accumulations in those external
regions. Carrying these back with continuity into the visible/HI
region, we find that the extent of the velocity dispersion as we
track curves at different non-zero $z$ values depends on the assumed
external velocity profile fall-off. (See, for example, the dispersion figure in
\cite{CT3}.)
With sufficient data, it should be possible, at least in 
principle, to provide limits on the extent of extra matter that might 
lie outside of the visible/HI region. To this point, we have only the data provided
in \cite{Fra} \cite{Bat} \cite{H} but far more data will be required to provide an
adequate discriminating test.

\section{Concluding Comments}

Readers are often confused as to how our results could be so
different from Newtonian predictions. They note that the dynamic solar system
analysis proceeds very accurately on the basis that the planets move in almost the
same manner using general relativity as that deduced by Newtonian dynamics and the
observations of the planetary motions confirm this.
However there is an essential difference between the solar system dynamics and the
galactic dynamics.
In the case of the solar system, the primary source of gravity is the sun and the 
planets are treated as \textit{test} particles in this field apart from
contributing minor perturbations when the slight changes are being sought. The
planets respond to 
the field of the sun but their own gravitational contributions are not retained
since they are so small.  By contrast, in the galaxy problem, the source of the
field 
is the combined rotating mass of all of the freely-gravitating 
elements themselves that compose the galaxy. There is no one single dominant
contributor in the galactic problem. The observed elements in the galactic case
\textit{produce} the field; they do not merely \textit{respond} to the field as they
do in the planetary problem.

We have seen that the non-linearity for the computation of density distribution
inherent in the Einstein field equations for a stationary 
axially-symmetric pressure-free mass distribution, even in the case 
of weak fields, leads to the incorporation of the known galactic velocity curves
with little or no extra dark matter. This is as 
opposed to the curves that had been derived on the basis of 
Newtonian gravitational theory that demanded the presence of vast extra reservoirs
of dark matter. It is worth repeating for emphasis: the results were consistent 
with the observations of velocity as a function of radius plotted as 
a rise followed by an essentially flat extended region and no halo of 
dark matter with multiples of the normally computed galactic 
mass was required to achieve them. The density distribution that is 
revealed thereby is one of an 
essentially flattened disk without an accompanying overwhelmingly
massive vastly extended dark matter halo. With the ``dark'' matter 
being associated with the disk which is itself visible, it is
natural to regard the non-luminous material as normal baryonic
matter.

To some extent we have had to assume extensions of matter distribution with
assumptions regarding ultimate velocity fall-offs beyond that which is actually
observed in order to make comparisons \cite{CT4}. In the course of these
investigations, 
we have seen that these can readily yield a picture of galactic 
structure without huge extended very massive halos of dark 
matter. It would be helpful if new data beyond those presently available would be
produced. This would help tie down the complete physical picture.

Of particular interest is that we have within our grasp a criterion 
for determining the extent, if of any significance, of extra matter 
beyond the visible and HI regions of a galaxy. It is possible in 
principle to determine this with data solely \textit{within} the 
visible/HI region by plotting the velocity dispersion of rotation 
curves for various $z$ values. This is an attractive area for future 
research. In particular, it expands the demands upon not only our galactic model but
also upon any other proposed model by other researchers. It asks for consistency
between observation and theoretical prediction for the overall averaged picture of
stellar motions within the galaxy.

We have seen with now six galaxies that Einstein's general relativity has within it
the power to reveal what is looking more and more like a general truth, that there
is a consistent average density for galaxies at the point where their visible light
first appears. This is a power that is lacking in the much simpler theory of gravity
of Newton.

It is to be noted that we have also analyzed an \textit{intrinsically dynamic}
free-fall model, that of an idealized Coma Cluster of galaxies \cite{CT5}. This work
has the following attractive features: 

a) It makes use of an \textit{exact} solution of the Einstein equations, hence
removing any issues arising from the application of approximations. 

b) Being free of any singularities of any kind during the weak-field period of
analysis, it eliminates any possibility of the kind of objections that we have
discussed in the present paper. 

c) It demonstrates that for weak fields and bodies with $V<<c$, Einstein's theory
leads to results that differ from those of Newton's theory. 

This work further reveals the greater richness of general relativity in dealing with
issues in galactic dynamics.

We share the belief of many that the scientific method has been most successful when
guided by 
Occam's razor, that new elements should not be introduced into a 
theory unless absolutely necessary. If dark matter should turn out to be another
case similar to the ether of the 19th Century, it is well for us to determine this
sooner rather than later.

\vskip 0.125in

{\bf Acknowledgment}
\vskip 0.125in 
 We are grateful to Steven Tieu for helpful discussions.



\end{document}